\documentclass[aps, prd, showpacs, superscriptaddress, ctexart, nofootinbib, twocolumn]{revtex4}

\usepackage{amssymb, amsmath, bm, dcolumn, epsf, graphicx, latexsym, slashed, simplewick}

\usepackage{color}

\def\be{\begin{equation}}
\def\ee{\end{equation}}
\def\bea{\begin{eqnarray}}
\def\eea{\end{eqnarray}}
\begin{document}

\title{Fermi-bounce cosmology and the fermion curvaton mechanism}

\author{Stephon Alexander}
\email{stephon.alexander@dartmouth.edu}
\affiliation{Center for Cosmic Origins and Department of Physics and Astronomy, Dartmouth College, Hanover, New Hampshire 03755, USA}

\author{Yi-Fu Cai}
\email{yifucai@physics.mcgill.ca}
\affiliation{Department of Physics, McGill University, Montr\'eal, QC, H3A 2T8, Canada}

\author{Antonino Marcian\`o}
\email{marciano@fudan.edu.cn}
\affiliation{Center for Field Theory and Particle Physics \& Department of Physics, Fudan University, 200433 Shanghai, China}

\noindent
\begin{abstract}
\noindent
A nonsingular bouncing cosmology can be achieved by introducing a fermion field with BCS condensation occurring at high energy scales. In this paper we are able to dilute the anisotropic stress near the bounce by means of releasing the gap energy density near the phase transition between the radiation and condensate states.  In order to explain the nearly scale-invariant CMB spectrum, another fermion field is required. We investigate one possible curvaton mechanism by involving one another fermion field without condensation where the mass is lighter than the background field. We show that, by virtue of the fermion curvaton mechanism, our model can satisfy the latest cosmological observations very well, and that the fermion species involved may realize a cosmological see-saw mechanism after one finely tunes model parameters.
\end{abstract}

\maketitle

\section{Introduction}

\noindent
Nonsingular bouncing cosmologies can resolve the initial singularity and horizon problems of the hot Big-Bang theory.  These types of cosmological scenarios appear in many theoretical constructions \cite{Calcagni:2009ar, Kiritsis:2009sh, Brandenberger:2009yt, Cai:2009in,Cai:2011tc, Poplawski:2011jz,Cai:2010zma, Cai:2011bs,Cai:2012ag,Biswas:2005qr, Biswas:2006bs,Cai:2007qw, Cai:2007zv, Cai:2008qw, Bhattacharya:2013ut,Buchbinder:2007ad, Creminelli:2007aq, Lin:2010pf,Brandenberger:2013zea,Martin:2003sf, Solomons:2001ef}. We refer to \cite{Novello:2008ra, Lehners:2008vx, Cai:2014bea} for recent reviews of various bouncing cosmologies. Most nonsingular bounce models are based on the matter fields with integer spins, {\it i.e.} the bosonic sector of the universe. However, the fundamental particles that make up the macroscopic world are dominated by fermion fields, and it is interesting to investigate their effects in the early universe.  Recently, it was found in \cite{Alexander:2008vt,Alexander:2014eva} that a nonsingular bounce can be achieved by means of a fermion field with a condensate state in the ultraviolet (UV) regime.

In this model, Einstein gravity is extended to have topological terms, which present gravitational interactions with the Dirac fermions, and torsion, which leads to Four-Fermion current densities and therefore effectively contribute to a negative energy density that evolves as $\sim a^6$.   In the infrared (IR) regime, the fermion-field is dominated by its mass term which generates a matter-like contraction preceding the nonsingular bounce. Thus, this model nicely supports the paradigm of the matter-bounce scenario \cite{Wands:1998yp, Finelli:2001sr} that is necessary for generating a nearly scale-invariant power spectrum of primordial perturbations in the contracting phase.

While the bounce model realized by this nontrivial cosmological fermion field can realize the matter-bounce scenario and provide a framework of generating power-spectrum of observational interest, it shares a common issue that exists in a large class of bouncing cosmologies. That is, their contracting phases are not stable against the instability to the growth of the anisotropic stress, which is known as the famous Belinsky-Khalatnikov-Lifshitz (BKL) instability issue \cite{Belinsky:1970ew}.  In the context of scalar field cosmology, this issue can be solved if there exists a phase with a steep and negative-valued potential which dominates over the anisotropies in the contracting phase \cite{Khoury:2001wf, Erickson:2003zm, Cai:2012va, Cai:2013vm}. A concrete realization of such a new matter-bounce by using scalar fields was constructed in \cite{Cai:2013kja} (see \cite{Koehn:2013upa, Cai:2014zga} for extended studies and \cite{Cai:2014bea} for a recent review).

In the present work we take a close look at the cosmological implication of a Dirac Fermion field and show that the new matter-bounce scenario can be achieved in this model due to the gap energy released during the transition from a regular massive fermion state to the state of a Four-Fermion condensation.  As the scale factor decreases, the vacuum expectation value of the fermion bilinear $\langle\bar\psi\psi\rangle$ grows as $\sim a^{-3}$ and evolves to a critical value that triggers the condensation phase transition. Then, the value of $\langle\bar\psi\psi\rangle$ is locked at the surface of a phase transition and therefore so is the scale factor.  As a result, a large amount of gap energy density which dominates anisotropic stress near the bounce.

Based on this scenario, we study a new curvaton mechanism by considering another flavor of a fermion field in which the mass is much lighter than the background one. Fluctuations of this fermion field, originating from a quantum vacuum state, can form a scale-invariant spectrum during the matter-like contracting phase. These fluctuations automatically dominate over the curvature perturbation at large length scales since those of the background field are suppressed by its mass term. Therefore, our two-flavor fermion field model provides a fermion curvaton mechanism for generating a scale-invariant power spectrum of primordial curvature perturbation which can explain the CMB temperature anisotropies \cite{Ade:2013zuv}. Moreover, by studying primordial tensor perturbations, we find that, by virtue of this mechanism, there exists a large parameter space for the tensor-to-scalar ratio to be consistent with the latest experiments, such as the BICEP2 data \cite{Ade:2014xna}.

The paper is organized as follows. In Section \ref{Sec:general}, we briefly review the bouncing cosmology by means of the fermion condensate. Then, in Section \ref{Sec:anisotropy} we present the detailed study on the theoretical constraint from primordial anisotropies. To address the instability issue arisen from primordial anisotropies, in Section \ref{Sec:model} we analyze the cosmological implication of the gap energy density restored in the fermion field and numerically show that this part of contribution can give rise to a period of contraction. Afterwards, in Section \ref{Sec:tensor} we study the primordial gravitational waves generated in this model. We conclude with a discussion in Section \ref{Sec:conclusion}. Throughout the paper we take the sign of the metric to be $(+,-,-,-)$ and define the reduced Planck mass through $\kappa\equiv 8\pi G =1/M_p^2$.

\section{The bounce cosmology with fermion condensate}\label{Sec:general}

\noindent
We start by briefly reviewing the cosmology of the fermion condensate model. Consider a universe filled with a fermion field, one can write down the action as
\be
S= S_{\rm GR} + S_{\rm \psi} \,,
\ee
where the Einstein-Hilbert action is expressed in terms of the mixed-indices Riemann tensor $R_{\mu\nu}^{IJ}=F_{\mu\nu}^{IJ}[\widetilde{\omega}(e)]$
\be
S_{GR}= \frac{1}{2 \kappa} \int_{M} \!\!\! d^4 x |e| e^\mu_I e^\nu_J R_{\mu\nu}^{IJ} \,,
\ee
the Dirac action $S_{\rm \psi}$ on curved space-time
reads
\be
S_{\rm \psi}= \frac{1}{2} \int_{M} \!\!\! d^4 x |e| \left( \
\overline{\psi} \gamma^I e^\mu_I \imath \widetilde{\nabla}_\mu \psi - m_\psi \overline{\psi} \psi \right) +{\rm h.c.}\,,
\ee
and finally the interacting part of the theory is:
\be \label{interact}
S_\psi^{\rm Int} \!=\! -\xi \kappa\! \int_{M} \!\!\! d^4 x |e| \, J_\psi^L\, J_\psi^M \, \eta_{LM}\,,
\ee
which only involve the axial vector current $J_\psi$ of the $\psi$ fermionic species. \\

Varying the action with respect to the vierbein can yield the following energy-momentum tensor
\be \label{tenfepsi}
\hspace{-0.25cm}
T^{\rm \psi}_{\mu\nu}\!=\! \frac{1}{4} \overline{\psi} \gamma_I e^I_{( \mu} \imath \widetilde{\nabla}_{\nu )} \psi +{\rm h.c.}  -  g_{\mu\nu} \mathcal{L}_{\rm \psi} \,.
\ee
Using comoving coordinates for a Friedmann-Lema\^itre-Robertson-Walker (FLRW) universe, we can solve the Euler-Lagrange equations of the system using the ansatz for the fermion field $\psi=(\psi_0,0,0,0)$, and find that
\be
\bar{\psi} \psi= \frac{n_\psi}{a^3} \,.
\ee
Using the Fierz identities, we can then write the first Friedmann equation taking into account the contributions due to the fermionic field:
\be
H^2 = \frac{\kappa}{3} m_\psi \, \frac{n_\psi}{a^3} + \xi\,\frac{\kappa^2}{3}\,\frac{n_{\psi}^2}{a^6} \,.
\ee
One can see the first term in the above expression corresponds to the regular phase of the fermion field which is dominated by the mass term. The second term appears when the fermion field enters the condensate state with $\xi<0$. When applied to the cosmological background, the fermion condensate state can effectively contribute to a negative energy density and hence cancels the energy densities from regular matter fields. Therefore, a nonsingular bouncing solution is achieved.

To be explicit, for a universe filled with a single fermion field $\psi$ as described above, one can derive the solution of the scale factor of the metric as:
\be
\label{adri}
 a\!=\!\!\left( \frac{3 }{4} \kappa m_\psi n_\psi (t-t_0)^2 \!-\! \frac{ \xi  \kappa \, n_\psi }{m_\psi }  \right)^{\frac{1}{3}}\!.
\ee
In this solution, one can see that the universe shrinks to the minimal size when $t=t_0$, {\it i.e.} $a_0 = (-\frac{\xi \kappa n_\psi }{m_\psi })^{1/3}$, which is real if $\xi<0$.

\section{Constraint from the anisotropic instability}\label{Sec:anisotropy}

\noindent
A general challenge for bouncing cosmologies is to ensure that the contracting phase is stable against the instability to the growth of anisotropic stress, whose energy density grows as $a^{-6}$. This is the famous BKL instability issue of any cosmological models involving a contracting phase. In particular, for the matter-bounce cosmology, the background energy density scales as $a^{-3}$ and hence, one needs to introduce a mechanism to suppress the growth of this unwanted anisotropies in the contracting phase. In the model of fermionic bounce, the energy density contributed by the fermionic condensate evolves also as $a^{-6}$ but with a negative sign. In this case, whether the model is stable depends on the initial condition one chooses, namely, it is marginally stable if initially the contribution of the anisotropic stress is much lower than that of the fermionic condensate. We perform an estimate on the theoretical constraint from the anisotropic instability.

In general, the anisotropic stress originate from the cosmic fluid through the decomposition of the energy stress tensor as follows,
\begin{align}
 T_{\mu\nu} = (\rho+P) u_\mu u_\nu +g_{\mu\nu} P +\pi_{\mu\nu}~,
\end{align}
where the $u_\mu$ is the 4-velocity, $\rho$ the energy density, $P$ the pressure, and $\pi$ the anisotropic stress tensor. The latter is related to the shear tensor $\sigma_{ij}$ through,
\begin{eqnarray}
 \dot\sigma^i_j +3 H\sigma^i_j = \frac{1}{M_p^2} ( \pi^i_j- \frac{1}{3}\delta^i_j \pi^k_k ) ~.
\end{eqnarray}
The existence of the shear tensor leads to a backreaction to the background energy density, contributing by
\begin{eqnarray}
 \rho_A = \frac{1}{2}\sigma^i_j \sigma^j_i ~.
\end{eqnarray}

By inserting the energy stress tensor into the above expression, one can easily estimate the contribution of the anisotropy in the model of fermionic bounce, which is roughly expressed as
\begin{eqnarray}
 \rho_A \sim \frac{{\rm Tr}(\gamma_i\gamma_j)}{M_p^2} \bar\psi\psi \langle\delta\bar\psi\delta\psi\rangle ~,
\end{eqnarray}
where we have taken the average spatial derivative of the fermion fluctuation in the order of the Hubble rate. From this expression, it is easy to see that the energy density from the anisotropic stress in the current model also evolves as $a^{-6}$ as $\psi\sim a^{-3/2}$. Then, to compare with the background equation of motion provided in the previous section, one can easily find that the model is marginally stable against the anisotropic instability only when
\begin{align}\label{constraint1}
 \langle \delta \bar\psi \delta \psi \rangle /M_p^2 \ll m ~.
\end{align}
In this case, the background universe will be of the FLRW form.

Following the analysis of Ref. \cite{Alexander:2014eva}, we have the relation between the power spectrum of curvature perturbation and the correlation function of the fermion fluctuations as
\begin{eqnarray}
 \mathcal{P}_\zeta \simeq O(1) \frac{\langle \delta \bar\psi \delta \psi \rangle}{4\bar\psi\psi} ~,
\end{eqnarray}
for the single fermion field model. Thus we can eliminate the correlation function in the constraint equation \eqref{constraint1}.  One can estimate the upper bound by making use of the maximal value of the $\bar\psi\psi$, which is the fermion density at the bounce $n_0$. Correspondingly, one derives the following constraint
\begin{eqnarray}
 m\gg 4\mathcal{P}_\zeta ~ n_0/M_p^2 ~,
\end{eqnarray}
for the mass of the fermion field. As a simple estimate, one may take $\mathcal{P}_\zeta\sim 10^{-9}$ and $n_0 \lesssim M_p^3$, and then the constraint becomes $m \gg 10^{-9} M_p$. However, one also needs to fine tune an extremely small value for the mass parameter at the very beginning of matter contraction so that the power spectrum generated in this stage is nearly scale invariant. To be specific, if we expect that primordial perturbations generated during matter contraction can cover at least 10 e-folds for the CMB sky, the corresponding mass is required to be $m\ll 10^{-30} n_0/M_p^2$. The constraint would be much more stringent if we consider the number of e-folds of primordial perturbations to be larger. As a result, it is difficult to reconcile the above two constraints consistently.

A solution of solving this anisotropy problem is to introduce a period of contraction in which the universe is dominated by a matter field with an equation of state parameter much larger than unity. To achieve this, one introduces a scalar field with a negatively valued exponential potential. However, in the model of fermionic bounce, we will find that the gap energy stored in the phase transition can realize this a large barytropic index, as will be shown in the next section.

\section{Fermion-bounce cosmology}\label{Sec:model}

\noindent
We consider in this section a variant of the model addressed in \cite{Alexander:2014eva}. We first take a closer look at the cosmic evolution during the phase transition which connects the matter contraction and the phase of the fermion condensate. We show that the gap energy stored in this stage can help to realize a period of fermionic contraction and hence, the unwanted anisotropy generated in matter contraction would be diluted out.  Furthermore, in order to meet with the constraint from the scale invariance of power spectrum, we make use of the curvaton mechanism by including one extra fermion field. The mechanism we are presenting works then as we were introducing a ``curvaton fermionic field'', instead of curvaton scalar field.

\subsection{Erasing anisotropic stress from the phase transition}

\noindent
In the previous study of the fermion bounce model \cite{Alexander:2014eva}, it is assumed that the gap energy is secondary during the phase transition. Along with the cosmic contraction, the value of the fermion bilinear increases and evolves to a critical value that is about to trigger the fermion condensate phase. If we take a closer look at the physical implication of the phase transition, it is interesting to observe that during this period, the gap energy can be released to increase the energy density of the universe and hence the amplitude of the Hubble rate. In the meanwhile, however, the value of the fermion bilinear $\bar\psi\psi$ does not necessarily vary dramatically. This implies, the scale factor of the universe can vary slowly during the phase transition and therefore a period of fermionic contraction is achieved. Correspondingly, we expect that the phase transition connecting the regular fermion phase and the fermion condensate phase can be used to give rise to the fermionic-matter bouncing solution.

\begin{figure}
\includegraphics[scale=0.4]{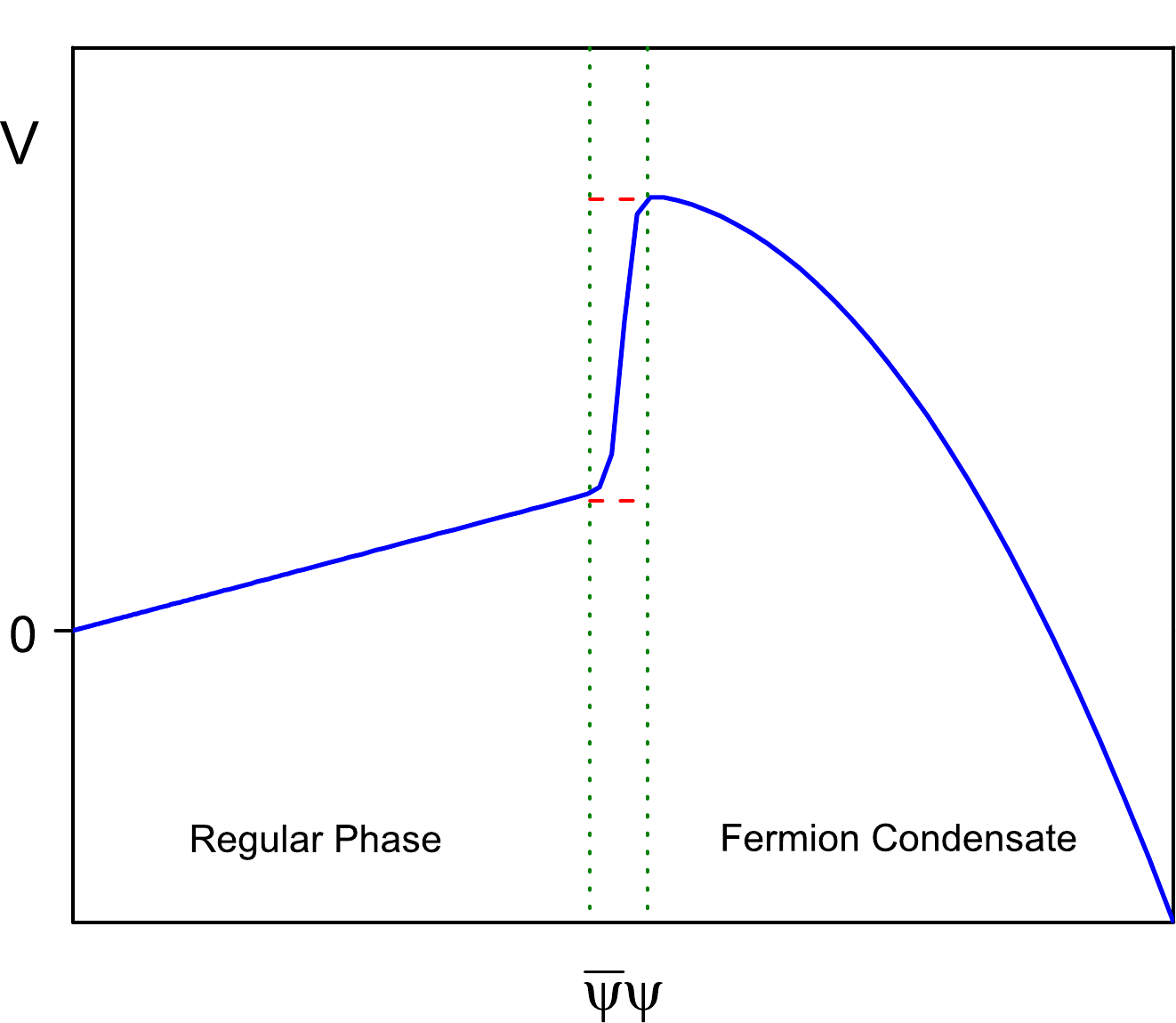}
\caption{A sketch of the potential V as a functions of the bilinear $\bar\psi\psi$. Along with the variation of $\bar\psi\psi$, the field space is separated into three regimes: the regular massive fermion, the fermion condensate, and the phase transition regime that connects the previous two. We find that the gap energy stored in the phase transition can effectively leads to a period of matter-bounce 
phase. }
\label{Fig:V}
\end{figure}

We depict the shape of the potential for the fermion field in the sketch plot in Fig. \ref{Fig:V}. In the figure, when $\bar\psi\psi$ lies in the left side of the green dot lines, the potential is a linear function and thus simply a mass term. In the right side of the green dot lines, the potential bends into a negative value due to the fermion condensate effect. The narrow regime separated by these two green dot lines corresponds to the phase transition stage. One can easily see that the gap energy can be released in this stage while the value of $\bar\psi\psi$ is almost conserved.
The potential for the fermion field with the above properties may be parameterized as the following form,
\begin{align}
 V &= m_\psi \bar\psi\psi \frac{1-\tanh[\alpha (\bar\psi\psi-(\bar\psi\psi)_*)]}{2}  \nonumber\\
 &+ [m_\psi(\bar\psi\psi)_* + \Delta\rho +\xi\kappa (\bar\psi\psi-(\bar\psi\psi)_*)^2 ] \nonumber\\
 & \times \frac{1+\tanh[\alpha (\bar\psi\psi-(\bar\psi\psi)_*)]}{2}
\end{align}
where $\alpha$ is introduced to smooth the slope of the potential near the phase transition and $\Delta\rho$ represents for the gap energy density. Additionally, $(\bar\psi\psi)_*$ denotes the value of the fermionic bilinear term around the phase transition.

\begin{figure}
\includegraphics[scale=0.35]{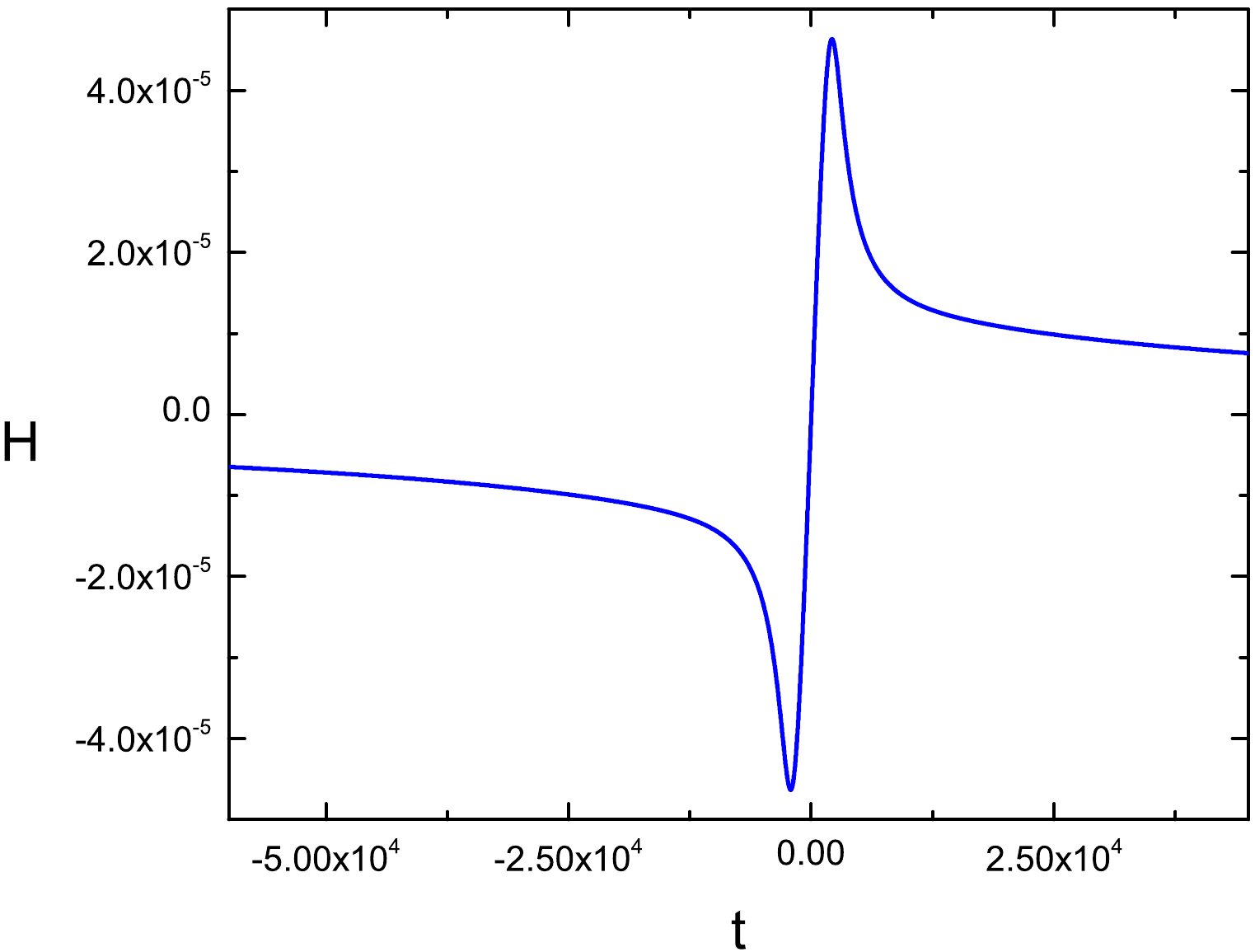}
\includegraphics[scale=0.35]{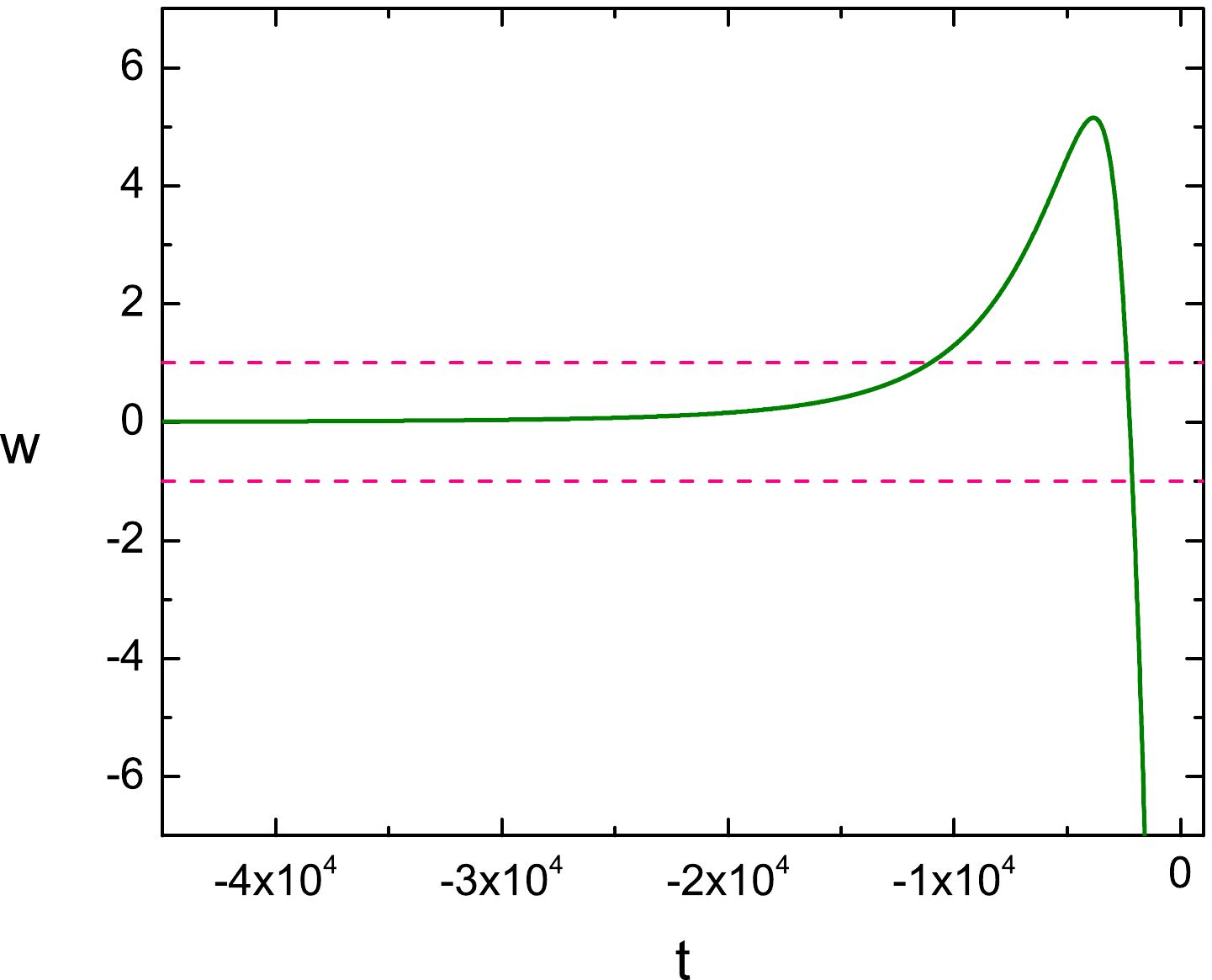}
\caption{Numerical plot of the evolutions of the Hubble parameter $H$ and the background EoS $w$ as a function of cosmic time in the model under consideration. In the numerical calculation, we take the values of model parameters as provided in the main text. All dimensional parameters are of Planck units. }
\label{Fig:Hw}
\end{figure}

In this section, we numerically examine the dynamics of the model under consideration. We show the evolutions of the Hubble parameter $H$ and the Equation of State (EoS) $w$ of the background universe in Fig. \ref{Fig:Hw}. In particular we choose the model parameters as follows: $m_\psi = 10^{-8} ~,~ \xi = -1\times10^{-5.5} ~,~ \alpha = 50 ~,~ (\bar\psi\psi)_* = 0.08 ~,~ \Delta\rho = 10^{-7.5} ~,~ m_\chi = 10^{-10}$, all dimensional parameters being of Planck units. We use the blue solid line to depict the Hubble parameter and the green solid line to represent for the EoS. From the evolution of the Hubble parameter, one can read that the universe transits from a contracting phase to an expanding one smoothly and when $t=0$, a nonsingular bounce takes place. The EoS $w$ initially equals zero, which implies that the universe behaves as a dust-like one during the contraction. Then $w$ evolves to above unity and correspondingly the background universe evolves into the matter-bounce phase for a while. At the bouncing point, $w$ decreases dramatically and evolves to negative infinity.

\begin{figure}
\includegraphics[scale=0.35]{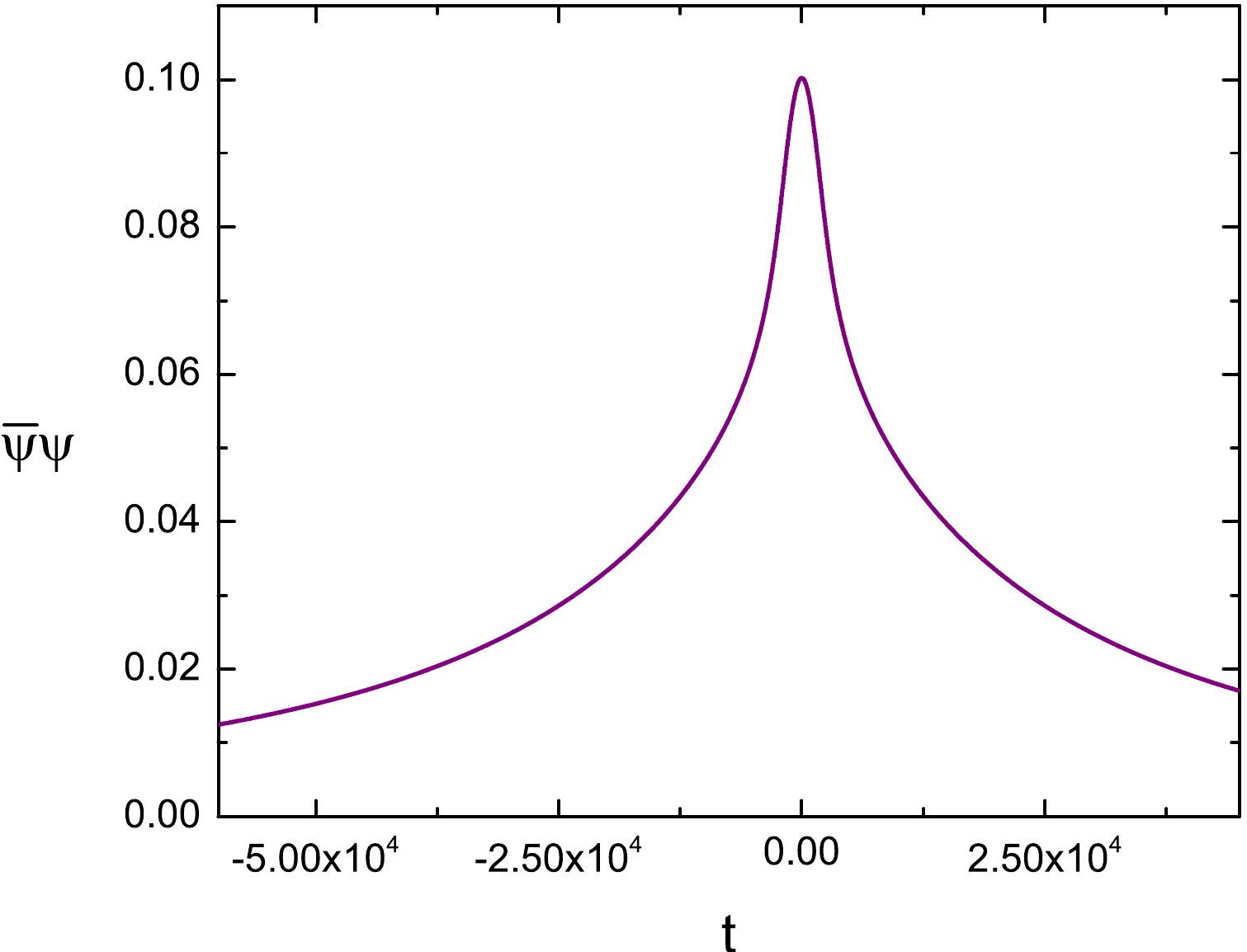}
\caption{Numerical plot of the evolution of the bilinear $\bar\psi\psi$ as a function of cosmic time in the model under consideration. In the numerical calculation, we take the values of model parameters as provided in the main text. All dimensional parameters are of Planck units. }
\label{Fig:app}
\end{figure}

Having known the dynamics of the Hubble parameter, one can integrate out the evolutions of the scale factor and then the scalar bilinear $\bar\psi\psi$ exactly. In order to better understand the background fermion field, we numerically track its dynamics throughout the whole evolution as shown in Fig. \ref{Fig:app}. From the figure, one can explicitly find that the bilinear $\bar\psi\psi$ reaches the maximal value at the bouncing point.

\subsection{A two-field model}
\noindent
In this section we show how to consistently generate a scale invariant scalar power spectrum with two fermion fields.  We start from the action
\be
S= S_{\rm GR} + S_{\rm \psi}+S_{\rm \chi}
+S_{\rm Int}\,,
\ee
where the Einstein-Hilbert action is expressed in terms of the mixed-indices Riemann tensor $R_{\mu\nu}^{IJ}=F_{\mu\nu}^{IJ}[\widetilde{\omega}(e)]$
\be
S_{GR}= \frac{1}{2 \kappa} \int_{M} \!\!\! d^4 x |e| e^\mu_I e^\nu_J R_{\mu\nu}^{IJ} \,,
\ee
the Dirac action $S_{\rm \psi}$ on curved space-time
reads
\be
S_{\rm \psi}= \frac{1}{2} \int_{M} \!\!\! d^4 x |e| \left( \
\overline{\psi} \gamma^I e^\mu_I \imath \widetilde{\nabla}_\mu \psi - m_\psi \overline{\psi} \psi \right) +{\rm h.c.}\,,
\ee
and finally the interacting part of the theory is:
\be \label{interact}
S_\psi^{\rm Int} \!=\! -\xi \kappa\! \int_{M} \!\!\! d^4 x |e| \,\left( J_\psi^L\, J_\psi^M + J_\chi^L\, J_\chi^M \right)\, \eta_{LM}\,,
\ee
which only involve the axial vector current $J_\psi$ of the $\psi$ fermionic species. \\

The two Dirac Lagrangians respectively reads,
\be
\mathcal{L}_{\rm \psi} \!= \!\frac{1}{2}\! \left( \overline{\psi} \gamma^I e^\mu_I \imath \widetilde{\nabla}_\mu \psi - m_\psi \overline{\psi} \psi \right) \!+ {\rm h.c.} -  \xi \kappa\, J_\psi^L J_\psi^K\, \eta_{LK}\, ,
\ee
and
\be
\!\!\mathcal{L}_{\rm \chi} = \frac{1}{2} \left( \overline{\chi} \gamma^I e^\mu_I \imath \widetilde{\nabla}_\mu \chi - m_\chi \overline{\chi} \chi \right) \!+ {\rm h.c.} -  \xi \kappa\, J_\chi^L J_\chi^K\, \eta_{LK}\,,
\ee
and yield the energy-momentum tensors
\be \label{tenfepsi}
\hspace{-0.25cm}
T^{\rm \psi}_{\mu\nu}\!=\! \frac{1}{4}  \overline{\psi} \gamma_I e^I_{( \mu} \imath \widetilde{\nabla}_{\nu )} \psi +{\rm h.c.}  -
 g_{\mu\nu} \mathcal{L}_{\rm \psi} \,,
\ee
and
\be \label{tenfechi}
\hspace{-0.25cm}
T^{\rm \chi}_{\mu\nu}\!=\! \frac{1}{4}  \overline{\chi} \gamma_I e^I_{( \mu} \imath \widetilde{\nabla}_{\nu )} \chi +{\rm h.c.}  -
 g_{\mu\nu} \mathcal{L}_{\rm \chi} \,.
\ee
We can solve the Euler-Lagrange equations of the system using the ans\"atze for the fermion fields $\psi=(\psi_0,0,0,0)$ and $\chi=(\chi_0,0,0,0)$, and find that
\be \label{psi_chi}
\bar{\psi} \psi= \frac{n_\psi}{a^3}\,, \qquad \bar{\chi} \chi= \frac{n_\chi}{a^3}\,.
\ee
Using the Fierz identities, we can then write the first Friedmann equation taking into account the contributions due to the two fermionic species
\be
H^2=\xi\,\frac{\kappa^2}{3}\,\frac{n_{\psi}^2}{a^6} + \frac{\kappa}{3} m_\psi \, \frac{n_\psi}{a^3}+ \xi\,\frac{\kappa^2}{3}\,\frac{n_{\chi}^2}{a^6} + \frac{\kappa}{3} m_\chi \, \frac{n_\chi}{a^3}\,.
\ee
The scale factor of the metric then reads
\be
\label{adri}
a\!=\!\!\left( \frac{3  \kappa (m_\psi n_\psi\!+\!m_\chi n_\chi ) }{4} (t-t_0)^2 \!-\! \frac{ \xi  \kappa \,( n_\psi^2 + n_\chi^2) }{(m_\psi n_\psi+m_\chi n_\chi)}  \right)^{\frac{1}{3}}\!,
\ee
therefore its value in $t_0$ it is found to be
\be \label{azero}
a_0\!=\!\!\left( - \frac{ \xi  \kappa \,( n_\psi^2 + n_\chi^2) }{m_\psi n_\psi+m_\chi n_\chi}  \right)^{\frac{1}{3}} \!\simeq\! \left( - \frac{ \xi  \kappa \,( n_\psi^2 + n_\chi^2) }{m_\psi n_\psi}  \right)^{\frac{1}{3}} \!.
\ee
We may now consider the perturbations to the energy density which is
\be \label{pro}
\zeta= \frac{\delta \rho}{\rho+p}\,.
\ee
Since we have two fermionic species with different values of the bare mass,
the two contributions to the variation of the energy densities read
\begin{eqnarray}
\!\!\!\! && \!\!\!\!\!\! \delta \rho= \\
\!\!\!\! && \!\! (m_\chi \!\!+\! \xi \kappa \overline{\chi} \chi ) \! \left(\delta\overline{\chi} \, \chi+ \overline{\chi} \, \delta \chi \right) \!+\! \left( m_\psi \!+\! \xi \kappa  \overline{\psi} \psi \right) \! \left(\delta\overline{\psi} \, \psi+ \overline{\psi} \, \delta \psi \right) \nonumber\\
\!\! && \!\!\!\! + \xi \kappa \Big[ \overline{\psi} \gamma_5 \psi ( \delta\overline{\psi} \, \gamma_5 \psi +\overline{\psi} \gamma_5 \delta \psi ) \!+\! \overline{\psi} \gamma^L \psi ( \delta\overline{\psi} \gamma_L \psi + \overline{\psi} \gamma_L \delta \psi )  \Big]
\nonumber\\
\!\! && \!\!\!\! + \xi \kappa \Big[ \overline{\chi} \gamma_5 \chi ( \delta\overline{\chi} \, \gamma_5 \chi +\overline{\chi} \gamma_5 \delta \chi ) \!+\! \overline{\chi} \gamma^L \chi ( \delta\overline{\chi} \gamma_L \chi + \overline{\chi} \gamma_L \delta \chi )  \Big]\,,
\nonumber
\end{eqnarray}
while for the denominator of (\ref{pro}) reads
\be
m_\chi \overline{\chi} \chi + m_\psi \overline{\psi} \psi + 2 \xi \kappa \left( J_\psi^L J_\psi^M +J_\chi^L J_\chi^M \right) \, \eta_{LM} \,.
\ee
Whenever we meet the requirement
\be \label{ansatz}
m_\psi n_\psi \gg m_\chi n_\chi \,,
\ee
we end up having for the $\zeta$ variable
\be
\zeta\simeq \frac{m_\chi  \, \left(\delta\overline{\chi} \, \chi+ \overline{\chi} \, \delta \chi \right)  }{m_\psi \overline{\psi} \psi } \,.
\ee
Therefore the autocorrelation function for $\zeta(t, \vec{x})$ now reads
\be \label{P_S_general}
 \mathcal{P}_S = \langle \zeta (t, \vec{x}) \zeta (t, \vec{x})  \rangle
  = \frac{m^2_\chi}{m^2_\psi} \frac{ \overline{\chi} \chi\, \langle \delta\overline{\chi} \delta\chi \rangle }{4(\overline{\psi} \psi)^2}  ~,
\ee
having assumed that: {\it i)} $m_\psi\!>\!\!>\!m_\chi$, so perturbations due to the $\psi$ field are suppressed at super-horizon scale, since their wavenumber are more blue-shifted with respect to the perturbations of the $\chi$ field; {\it ii)} cross-correlation between perturbations of the two different fermionic species can be neglected, since they are due to an interaction involving a graviton loop, which suppressed by the forth power of the Planck mass $M_p$.

The perturbations to the $\chi$ field can now be computed resorting to the same kind of assumptions discussed in \cite{Alexander:2014eva}. The scale factor for the metric, away from the bounce, reads again $a(\eta) \simeq \eta^2/\eta_0^2$, but now
\be \label{eta}
\eta_0=[\kappa (m_\psi n_\psi + m_\chi n_\chi) ]^{-1/2}\,,
\ee\\
which in the assumption (\ref{ansatz}) becomes
\be \label{eta}
\eta_0\simeq  (\kappa m_\psi n_\psi )^{-1/2}\,.
\ee
The equation for the perturbations of the $\chi$ field now reads
\be
\left( \gamma^I e_I^\mu \imath \widetilde{\nabla}_\mu - m_\chi -2 \xi\kappa \overline{\chi}_g \chi_g \right) \delta\chi=0\,,
\ee
in which we may use the background solution $\chi_{\rm g}$ so far recovered. We then reshuffle the equation of motion for the spinor perturbations and densitize them
\be
\left( \gamma^I e_I^\mu \imath \widetilde{\nabla}_\mu - m_\chi - 2 \xi\kappa \sqrt{-g} \,\widetilde{\overline{\chi}}_g \widetilde{\chi}_g \right) \widetilde{\delta \chi}=0\,.
\ee
Using the background $\chi$-fermion density, the latter equations recasts as
\be \label{Diri}
\left( \imath \gamma^\mu \partial_\mu - m_\chi \, a(\eta) -\frac{2 \xi \kappa \, n_\chi}{a^2(\eta)} \right) \widetilde{\delta\chi}=0 \, .
\ee

Following the procedure in \cite{Alexander:2014eva}, we can solve the Dirac equation (\ref{Diri}) in terms of
\begin{eqnarray}
&& \tilde{f}_{\pm h}= \frac{1}{\sqrt{2}} [\tilde{u}_{L,h} (\vec{k}, \eta)+\tilde{u}_{R,h} (\vec{k}, \eta)] \,,\nonumber\\
&&  \tilde{g}_{\pm h}=  \frac{1}{\sqrt{2}} [\tilde{v}_{L,h} (\vec{k}, \eta)+\tilde{v}_{R,h} (\vec{k}, \eta)] .
\end{eqnarray}

These come from rescaling densitized spinors up to $\tilde{u}=a^{3/2} u$ and $\tilde{v}=a^{3/2} v$, in terms of their chiral and helical components
\begin{eqnarray}
&& \hspace{-0.8cm}
\tilde{u}(t, \vec{k}) = \sum_h \tilde{u}_h(t, \vec{k})= \sum_h \left(
\begin{array}{c} \tilde{u}_{L, h} (\vec{k}, \eta) \\ \tilde{u}_{R,h} (\vec{k}, \eta)  \end{array}\right) \xi_h\,, \\
&& \hspace{-0.8cm}
\tilde{v}(t, \vec{k}) = \sum_h \tilde{v}_h(t, \vec{k})= \sum_h \left(
\begin{array}{c} \tilde{v}_{R, h} (\vec{k}, \eta) \\ \tilde{v}_{L,h} (\vec{k}, \eta)  \end{array}\right) \xi_h\,,
\end{eqnarray}
having introduced the helicity $2$-eigenspinor, cast in terms of the unit vector $\hat{\vec{k}}$, which reads
\be
\xi_h\!=\! \frac{1}{\sqrt{2(1-h\,\hat{k}_z)}}\!\left(\begin{array}{c}  h(\hat{k}_x - \imath \hat{k}_y) \\ \imath \hat{k}_x -h\,  \hat{k}_y \end{array}\right)\!, \ \ \hat{\vec{k}}\!\cdot\! \vec{\sigma}\, \xi_h=h \,\xi_h,
\ee
$\vec{\sigma}$ denoting the Pauli matrices.

In terms of $\tilde{f}_h$, we rewrite equation (\ref{Diri}) as
\be \label{completa}
\tilde{f}''_{\pm h} + \omega^2(k,\eta) \tilde{f}_{\pm h}=0 ~,
\ee
with an effective frequency term being defined by
\begin{eqnarray}\label{frequency}
 \omega^2(k,\eta) = k^2 \!+\! m^2_\chi a^2\!+\! \imath m_\chi a'
+ 2 \xi \kappa n_\chi \!\left(\frac{m_\chi}{a} \!-\! \imath \frac{a'}{a^3}\right)\!.
\end{eqnarray}
Since in our model $\chi$ plays the role of a curvaton which does not contribute to the background evolution, the condition $m_\chi \! \ll \! m_\psi$ holds and then one can neglect the second term of the effective frequency. In addition, the third and the last term are imaginary and thus can be smoothed out by taking the time-averaged evolution. Finally, the effective frequency mainly depends on the gradient term $k^2$ and the effective mass term $2 \xi \kappa n_\chi m_\chi/a$.

Afterwards, we can solve the above solution in two limits. First, we consider the gradient term to be dominant, which corresponds to the sub-Hubble scales with $|k\eta| \gg 1$. In this limit, it is natural to impose the initial condition for the perturbation modes by virtue of a Wentzel-Kramers-Brillouin approximation, which then yields
\begin{eqnarray}\label{tilde_f_in}
 \tilde{f}_{\pm h} \simeq \sqrt{\frac{m_\chi}{\!\!2 k}}  e^{-ik\eta} ~.
\end{eqnarray}
This initial condition exactly coincides with the vacuum fluctuations. Second, we study the asymptotic solution to the perturbation equation in the limit of $|k\eta| \ll 1$, {\it i.e.} at super-Hubble scales. To apply the relation  $a(\eta) \simeq \eta^2/\eta_0^2$ and Eq. \eqref{eta}, one can write down the effective mass term as
\begin{eqnarray}
 -\frac{\gamma}{\eta^2} ~~ {\rm with} ~~ \gamma = -\frac{2\xi n_\chi m_\chi}{n_\psi m_\psi} ~.
\end{eqnarray}
Then, the equation of motion yields another asymptotic solution, of which the leading term takes the form
\begin{eqnarray}\label{tilde_f_out}
 \tilde{f}_{\pm h} \simeq c(k)\, \eta^{\frac{2(1+\gamma-\sqrt{1+4\gamma})}{3-\sqrt{1+4\gamma}}}~,
\end{eqnarray}
where $c(k)$ is a $k$ dependent coefficient to be determined by matching the above two asymptotic solutions \eqref{tilde_f_in} and \eqref{tilde_f_out} at the moment of Hubble crossing. As a result, the asymptotic solution at super-Hubble scales is given by
\begin{eqnarray}\label{tilde_f_sol}
 \tilde{f}_{\pm h} \simeq \sqrt{\frac{m_\chi}{\!\!2 k}} (k\eta)^{\frac{2(1+\gamma-\sqrt{1+4\gamma})}{3-\sqrt{1+4\gamma}}} ~.
\end{eqnarray}

Substituting \eqref{tilde_f_sol} into the expression \eqref{P_S_general} yields the power spectrum of primordial curvature perturbations as follows,
\begin{eqnarray}\label{P_S_sol}
 \mathcal{P}_S = \frac{m_\chi^3 n_\chi }{m_\psi^2 n^2_\psi} \frac{k^2}{4\pi^2 a^2 } (k\eta)^{\frac{4(1+\gamma-\sqrt{1+4\gamma})}{3-\sqrt{1+4\gamma}}} ~,
\end{eqnarray}
where we have applied the relations in \eqref{psi_chi}. It is interesting to notice that, when $\gamma = 2$ ({\it i.e.} $\xi = -n_\psi m_\psi/n_\chi m_\chi$), the above power spectrum is exactly scale invariant and the corresponding amplitude scales as $\eta^{-2}$ during the matter contracting phase. Therefore, the scale invariant power spectrum generated in the fermion curvaton mechanism is expresses as
\begin{eqnarray}\label{P_S_matter}
 \mathcal{P}_S = \frac{m_\chi^3 n_\chi }{m_\psi^2 n^2_\psi} \frac{1}{4\pi^2 a^2 \eta^2} ~,
\end{eqnarray}
during the matter contracting phase.

We eventually evaluate the above expression at the end of the matter contracting phase $t_E$, at which the scale factor takes the value $a_E$, which then becomes,
\begin{eqnarray}\label{P_S_final}
 \mathcal{P}_S = \frac{m_\chi^3 n_\chi}{m_\psi^2 n^2_\psi} \frac{{H}_E^2}{16\pi^2} ~,
\end{eqnarray}
where $\eta_E = 2/{\cal H}_E=2/(a_E H_E)$ has been applied. Notice that the time $t_E$ is also the beginning of the phase transition, at which perturbations become constant already, throughout the rest of the primordial epoch, until these reenter the Hubble horizon.

If $\gamma$ slightly deviates from $2$ in (\ref{P_S_sol}), one can derive the following expression for the spectral index
\begin{eqnarray}
 n_S - 1 \equiv \frac{d\ln \mathcal{P}_S}{d\ln k} \simeq -\frac{2}{3}(\gamma-2) ~,
\end{eqnarray}
which accounts for the spectrum to be red-tilted.

\iffalse

The conditions to be fulfilled in order to reduce (\ref{completa}) to
\be \label{ino}
\!\!\tilde{f}''_{\pm h} \!\!+\!\! \left(k^2 \!-\! \frac{\nu^2-1}{4\, \eta^2} \! \right) \!\tilde{f}_{\pm h}\!=\!0\,,
\ee
whose solutions lead to a scale-invariant power-spectrum, now read
\be
m_\chi^2 m_\psi n_\psi \kappa^2\!<\!\!<\!1\,,
\ee
and
\be
\frac{n_\chi}{n_\psi^2} \! < \!\! < \! \kappa^{\frac{5}{2}} m_\psi^2\,.
\ee
Notice also that, differently from \cite{Alexander:2014eva}, we now find
\be \label{scin}
\nu^2 = 1 - 8 \, \xi \, \frac{(m_\chi n_\chi)}{(m_\psi n_\psi)}\, \frac{\kappa}{\eta_0^2}\,,
\ee
which allows more realistic values of the masses of the species involved, by allowing large variation for $\xi$. \\

Finally, the power-spectrum we recover for the adiabatic scalar perturbations reads
\be
\mathcal{P}_S= \frac{m^2_\chi}{m^2_\psi} \frac{k^3 |\Gamma(\nu)|^2}{8 \pi^2 n_\psi}  (- k \eta )^{-2 |\nu|}\,,
\ee
which is scale invariant for a choice of $\nu=3/2$, giving
\be\label{PS_curvaton}
\mathcal{P}_S= \frac{1}{32 \pi n_\psi \eta_0^3} \frac{m^2_\chi}{m^2_\psi} \,,
\ee

\fi

\section{Predictions on primordial gravitational waves}\label{Sec:tensor}

\noindent
In this section, we perform a calculation of the power spectrum of primordial gravitational waves in the present model. Note that, the evolution of primordial gravitational waves decouple from other perturbation modes at linear order and depends only on the background dynamics. Since our model provides an explicit realization of the new matter-bounce scenario, we can directly follow the detailed calculation in \cite{Cai:2013kja} (see also \cite{Cai:2014xxa}) and directly write down,
\begin{eqnarray}
 \mathcal{P}_T = \frac{1}{\vartheta^2}\frac{{\cal H}_E^2}{a_E^2 M_p^2} ~,~~ {\rm with}~~ \vartheta = 8\pi(2q-3)(1-3q)~,
\end{eqnarray}
where ${\cal H}_E$ and $a_E$ respectively denote the values of the comoving Hubble parameter and of the scale factor at the end of matter contracting phase, just right before the phase transition. The coefficient $q$ is a background parameter associated with the contracting phase, and thus in our model is determined by the detailed procedure of the phase transition, which typically is required to be less than unity.

From the perspective of theoretical consideration, if the universe evolves through the bounce without a fermionic contraction phase, the maximal amplitude of the Hubble rate is of order $\frac{m_\psi}{\sqrt{\xi}}$. Thus, in our model, when the fermionic matter-bounce phase occurs, the amplitude of the corresponding Hubble parameter has to be at most the same order of $\frac{m_\psi}{\sqrt{\xi}}$. As an approximation, we can estimate the maximal case by suggesting, $|H_E| \simeq \frac{m_\psi}{\sqrt{\xi}}$, and therefore, the corresponding power spectrum is approximately given by
\begin{eqnarray}
 \mathcal{P}_T \simeq \frac{1}{\vartheta^2}\frac{m_\psi^2}{|\xi| M_p^2} ~.
\end{eqnarray}
Accordingly, following the definition $r\equiv \mathcal{P}_T/\mathcal{P}_S$, one can write down the tensor-to-scalar ratio of our model as follows,
\begin{eqnarray}\label{r_tts}
 r = \frac{16\pi^2}{\vartheta^2}
 \frac{m_\psi^2}{m_\chi^3} \frac{n^2_\psi}{n_\chi M_p^2} ~,
\end{eqnarray}
where we have approximately taken the condition of scale invariance $\gamma\simeq 2$.

The latest cosmological observations indicate that $\mathcal{P}_S \simeq 2.2\times 10^{-9}$ \cite{Ade:2013zuv}. While for the tensor-to-scalar ratio the value $r \simeq 0.2$ detected by BICEP2 \cite{Ade:2014xna} has been recently questioned in \cite{Mortonson:2014bja} and \cite{Flauger:2014qra}, in which it has been shown that dust could still account for all or most part of the signal of the primordial gravitational waves.

We disregard the hypothesis of non-detection of gravitational waves in this analysis, and take into account a non-vanishing value for $r$ consistent with the error bars, namely $r\sim 10^{-2}$. The values of $\mathcal{P}_S$ and $r$ so far discussed, one applied into Eqs. \eqref{P_S_final} and \eqref{r_tts}, require two further constraints. The first one turns out to be a constraint on the mass of the heavy species:
\be \label{cosmapsi}
 m_\psi^2 \lesssim 10^{-11} \, |\xi|\, M_p^2~.
\ee
The assumption (\ref{ansatz}) implies large values of $|\xi|=(n_\psi m_\psi)/ (n_\chi m_\chi)$, once in presence of a nearly scale-invariant power spectrum ($\gamma\simeq2$). The new constraint (\ref{cosmapsi}) can then be thought as linking the mass of the heavy species to the GUT scale, if a proper choice of $\xi\simeq 10^{4}$ is done.

Combining Eqs. \eqref{P_S_final} and \eqref{r_tts}, we derive a further constraint
\be
 \frac{m_\psi^2}{m_\chi^3 } \frac{n_\psi^2}{n_\chi M_p^2} \sim O(1)~. \label{cosmachi}
\ee
Once suitable values of the femionic species' densities have been fixed, the latter constraint can be shown to be achieved in this model, and to link the mass of the light species to the mass of the heavy one.

\section{Conclusion}\label{Sec:conclusion}

\noindent
To conclude, in the present paper we have studied a nonsingular matter-bounce universe, which has been achieved by introducing a background fermion field with a condensation occurring at high energy scales. In the literature, it was already found that, embedding a nonconventional spinor field into the FLRW universe, one can derive a wide class of cosmological solutions. A spinor field was indeed applied in the study of inflationary cosmology \cite{ArmendarizPicon:2003qk,Magueijo:2012ug}, dark energy models \cite{Cai:2008gk, Alexander:2009uu}, and the emergent universe scenario \cite{Cai:2012yf, Cai:2013rna}. Bearing in mind the purpose of realizing a concrete example of spinor matter-bounce cosmology that could be consistent with the latest cosmological data, we have shown how the gap energy density, eventually restored in the regular state of a cosmic fermion, can yield a short period of ekpyrotic phase during the contracting phase of the universe. The derivation of a nearly scale-invariant CMB power-spectrum requires the introduction of another matter field. As in the ekpyrotic scenario, primordial anisotropies can be washed out during the universe's contraction, if we consider a curvaton mechanism that involves one another fermion field (without condensation), of which the mass is lighter than the background field. Anisotropies can be neglected when the amplitude of the curvaton perturbations, to which they are associated, is found to be much smaller than the mass of the background spinor. The contribution from the latter particle to energy density perturbations can be indeed neglected, because of the mass hierarchy with curvaton spinor. The fermion curvaton mechanism here developed enables this model to be consistent with the latest cosmological data. Furthermore, it bestows a framework in which realizing a see-saw mechanism endowed with phenomenological consequences in cosmology. Assuming for the relative abundance of the heavy $\psi$ fermions over the light $\chi$ fermions the value
\be \label{gora}
\frac{n_\psi}{n_\chi} \simeq 10^{7} ~ \frac{m_\chi^3}{n_\psi}\,,
\ee
might allow in this context the see-saw mechanism. Indeed, it would encode as the two fermions: {\it i)} a regular neutrino, accounting for the light $\chi$ species of this model, with a mass $m_\chi<10^{-3} eV$ that fulfills constraints from Big-Bang nucleosynthesis and it is still compatible with experimental data \cite{Battye:2013xqa}; {\it ii)} a sterile neutrino, which corresponds to the background field {\it i.e.} the $\psi$ species, with mass at the GUT scale for the choice of the ratio in (\ref{gora}). Smaller values than the GUT scale for the mass of the background species $\psi$ may be easily accounted in this model.

\begin{acknowledgments}
\noindent
We thank Robert Brandenberger, Robert Caldwell and Leonardo Modesto for useful discussions. The work of CYF is supported in part by Department of Physics in McGill University. AM acknowledges hospitality at Dartmouth College, where the project has started, and support from a start-up grant provided by Fudan University.
\end{acknowledgments}

\end{document}